\documentclass[twocolumn,showpacs,preprintnumbers,amsmath,amssymb,superscriptaddress]{revtex4}

\usepackage[pdftex]{graphicx}
\usepackage[pdftex,colorlinks=true,urlcolor=blue,linkcolor=black]{hyperref} 

\usepackage{dcolumn}
\usepackage{bm}

\begin{document}


\title{Atom-Dimer Scattering in a Three-Component Fermi Gas} 


\author{T. Lompe}
\email{thomas.lompe@mpi-hd.mpg.de}
\author{T. B. Ottenstein}
\author{F. Serwane}
\affiliation{Physikalisches Institut, Ruprecht-Karls-Universit\"at Heidelberg, Germany}
\affiliation{Max-Planck-Institut f\" ur Kernphysik, Saupfercheckweg 1, 69117 Heidelberg, Germany}
\affiliation{ExtreMe Matter Institute EMMI, GSI Helmholtzzentrum f\"ur Schwerionenforschung, Darmstadt, Germany}
\author{K. Viering}
\affiliation{Department of Physics, The University of Texas at Austin, Austin, TX 78712}
\author{A. N. Wenz}
\author{G. Z\"urn}
\affiliation{Physikalisches Institut, Ruprecht-Karls-Universit\"at Heidelberg, Germany}
\affiliation{Max-Planck-Institut f\" ur Kernphysik, Saupfercheckweg 1, 69117 Heidelberg, Germany}
\author{S. Jochim}
\affiliation{Physikalisches Institut, Ruprecht-Karls-Universit\"at Heidelberg, Germany}
\affiliation{Max-Planck-Institut f\" ur Kernphysik, Saupfercheckweg 1, 69117 Heidelberg, Germany}

%

\date{\today}

\begin{abstract}
Ultracold gases of three distinguishable particles with large scattering lengths are expected to show rich few-body physics related to the Efimov effect. We have created three different mixtures of ultracold $^6$Li atoms and weakly bound $^6$Li$_2$ dimers consisting of atoms in three different hyperfine states and studied their inelastic decay via atom-dimer collisions.
We have found resonant enhancement of 
the decay due to the crossing of Efimov-like trimer states with the atom-dimer continuum in one mixture as well as minima of the decay in another mixture, which we interpret as a suppression of exchange reactions of the type $\left|12\right\rangle+\left|3\right\rangle \rightarrow \left|23\right\rangle+\left|1\right\rangle$. Such a suppression is caused by interference between different decay paths and demonstrates the possibility to use Efimov physics to control the rate constants for molecular exchange reactions in the ultracold regime. 

\end{abstract}

\maketitle

 The quantum-mechanical three-body problem is a fundamental, yet challenging problem of few-body physics. In 1970 \mbox{V. Efimov} studied three-body systems of identical bosons whose two-body interactions can be described completely by the s-wave scattering length $a$. He found 
that for diverging scattering length such a system exhibits an infinite series of universal three-body bound states, which are called Efimov states \cite{efimov}. For negative scattering lengths the binding energies of these Efimov states become zero at critical values of the scattering length spaced by a universal scaling factor $e^{ \pi /s_0} \approx 22.7$, where $s_0 \approx 1.00624$ is a universal scaling parameter. For positive scattering length there is a weakly bound dimer state, and the Efimov trimers disappear when they cross the atom-dimer threshold, i. e. their binding energy becomes degenerate with the binding energy of a weakly bound dimer. The positions of the crossings are fixed by a single three-body parameter, which describes the effects of short-range three-body physics \cite{braaten_review}.

Ultracold gases are perfectly suited to study such physics, as they provide systems where the s-wave scattering length 
can be much larger than the range $r_0$ of the interatomic potential and then describes the two-body interactions almost perfectly.
Using Feshbach resonances \cite{feshbach_review} it is possible to tune the scattering length by applying a homogeneous magnetic field. For negative scattering length one can tune the scattering length to values where a trimer state crosses into the three-atom continuum. This leads to a zero-energy resonance in the three-atom scattering, which can be observed through enhanced three-body recombination into deeply bound dimers \cite{innsbruck_efimov}. For positive scattering lengths one can observe these trimer states either by looking for interference minima in three-body recombination \cite{innsbruck_efimov,lens_homo,lev_li7,randy_li7}, or more directly through observing resonant enhancement of atom-dimer relaxation at the crossing of a trimer state and the atom-dimer threshold \cite{innsbruck_atomdimer}. 
Recent studies have begun to go beyond the basic Efimov scenario, studying for example universal four-body states \cite{greene_4body,innsbruck_4body} or heteronuclear systems \cite{lens_hetero}. 

\begin{figure} [tb]
\centering
	\includegraphics [width= 8cm] {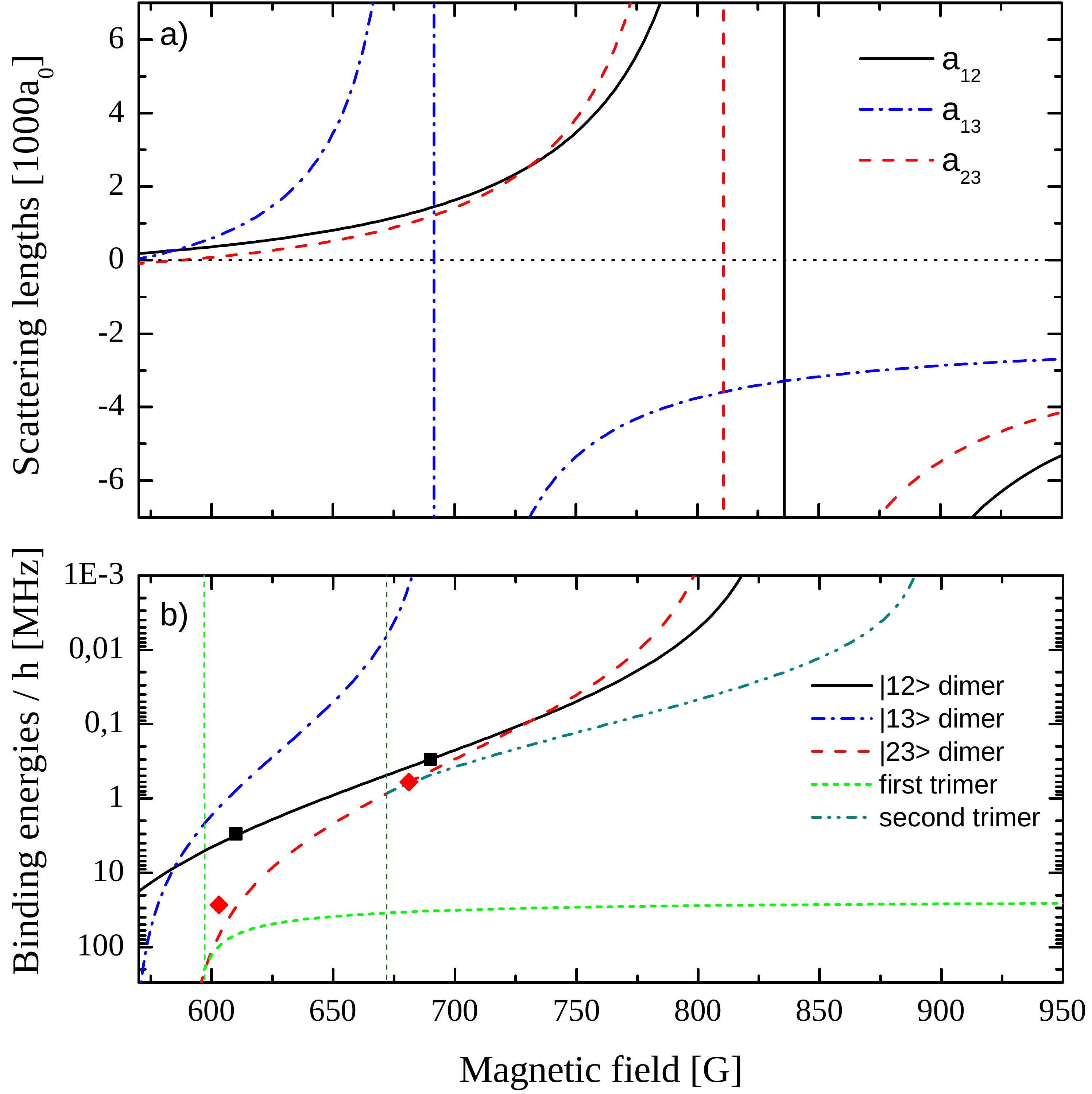}
	\caption{(a) Magnetic field dependence of the two-body s-wave scattering lengths for different channels in units of Bohr's radius $a_0$ \cite{julienne}. (b) Binding energies of the universal dimer and trimer states according to \cite{braaten_6li}. The vertical lines mark the predicted crossings of the trimer states with the $\left|1\right\rangle$-$\left|23\right\rangle$ atom-dimer threshold. The red diamonds mark the positions of the resonant enhancement observed in $\beta_{1-23}$ and the black squares mark the interference minima in $\beta_{3-12}$. }
	\label{fig:a_and_eb}
\end{figure}

In this Letter we study the case of three distinguishable fermions with large interparticle scattering lengths. In this case the interactions are described by three different scattering lengths, and consequently there can be more than one weakly bound dimer state in the system (see fig. \ref{fig:a_and_eb}). This allows to study phenomena which are not present for the case of identical bosons, as for example the effect of Efimov states on molecular exchange reactions.
For our studies we use fermionic $^6$Li atoms in the three lowest-energy hyperfine states, which we label $\left|1\right\rangle$, $\left|2\right\rangle$ and $\left|3\right\rangle$ \cite{wir_unten}.
As $^6$Li atoms are fermions only non-identical particles interact via s-wave scattering. These interactions are described by the three scattering lengths $a_{12}$, $a_{23}$ and $a_{13}$ for the respective combinations (see fig. \ref{fig:a_and_eb} a)). $^6$Li has the unique advantage that there are broad, overlapping Feshbach resonances for all these combinations, which allows to simultaneously tune all three scattering lengths by applying an external magnetic field. For each of these resonances there is a weakly bound dimer state with binding energy $E_{ij} = \hbar^2/m a_{ij}^2$, where m is the mass of a $^6$Li atom and $\hbar$ is the reduced Planck constant. We designate these dimer states $\left|12\right\rangle$, $\left|23\right\rangle$ and $\left|13\right\rangle$ respectively (see fig. \ref{fig:a_and_eb} b)). 
The first experiments with three-component gases of $^6$Li found two three-body loss resonances in the low-field universal region ($100\,$G - $500\,$G) \cite{wir_unten,ohara_unten}, which were subsequently identified as signatures of an Efimov state \cite{wetterich_ourdata,braaten_ourdata,naidon_ourdata,wenz_trimer}.
Another three-body loss resonance was later observed above the Feshbach resonances at 895\,G \cite{o'hara_oben}. This resonance belongs to the second Efimov trimer in the series. The first Efimov trimer never becomes unbound in the high-field region due to the large background scattering length.
Using the three-body parameter determined from the resonance at 895\,G, E. Braaten et al. calculated the spectrum of Efimov trimers across the resonances \cite{braaten_6li}. As the three scattering lengths do not diverge at the same magnetic field, the series of Efimov trimers is finite; according to universal theory there are two trimer states, whose energies are expected to become degenerate with the energy of a $\left|23\right\rangle$ dimer and a free atom in state  $\left|1\right\rangle$ at $672\,$G and $597\,$G. One should note that $a_{23} < r_0 \approx 60\, a_0$ at $597\,$G and the universal prediction is therefore not expected to be quantitatively accurate in this case. 

Here we study all possible combinations of free atoms in one state and dimers consisting of atoms in the two other states, which we call the $\left|1\right\rangle$-$\left|23\right\rangle$, $\left|3\right\rangle$-$\left|12\right\rangle$ and $\left|2\right\rangle$-$\left|13\right\rangle$
 mixtures.
For the $\left|1\right\rangle$-$\left|23\right\rangle$ mixture there is a zero-energy resonance in the $\left|1\right\rangle$-$\left|23\right\rangle$ atom-dimer scattering when the binding energy of the trimer state becomes degenerate with the binding energy of the $\left|23\right\rangle$ dimer, which leads to enhanced relaxation of weakly bound $\left|23\right\rangle$-dimers into deeply bound states. As the energy released in these processes is much larger than the trap depth, these events lead to the loss of one atom and one dimer from the trap. Therefore, this mixture should exhibit a resonant enhancement of two-body loss at these magnetic field values.

To create such a mixture we first prepare a thermal gas with twice as many atoms in state $\left|1\right\rangle$ as in state  $\left|2\right\rangle$ in an optical dipole trap, using the same scheme as described in \cite{wir_unten} to prepare the sample close to the zero-crossings of the scattering lengths.
 Due to the weak interactions we can then easily transfer the atoms in state $\left|2\right\rangle$ to state $\left|3\right\rangle$ with a rapid adiabatic passage, followed by a second passage transferring the atoms from state $\left|1\right\rangle$ to state $\left|2\right\rangle$. Thus, we create an imbalanced $\left|2\right\rangle$-$\left|3\right\rangle$ mixture with about $10^5$ atoms in state $\left|2\right\rangle$ and $5 \times 10^4$ atoms in state $\left|3\right\rangle$.
We then quickly change the magnetic field to $800\,$G, just below the $\left|2\right\rangle$-$\left|3\right\rangle$ Feshbach resonance and reduce the trap depth by a factor of two in a final stage of evaporative cooling. Additionally, any atoms left in state $\left|1\right\rangle$ are lost in three-body collisions during this stage. Next, we ramp the magnetic field down to $740\,$G, making sure that the ramp is slow enough that the sample stays in chemical equilibrium until almost all atoms in state $\left|3\right\rangle$ are bound in $\left|23\right\rangle$ molecules \cite{cheng_equilibrium,note_free_minority}, giving us roughly equal amounts of state $\left|2\right\rangle$ atoms and $\left|23\right\rangle$ molecules. As the radio-frequency transitions are shifted for the atoms bound in molecules it is possible to create a three-component sample from this mixture by selectively driving the free atoms in state $\left|2\right\rangle$ to state $\left|1\right\rangle$ with a rapid adiabatic passage. However, this requires the sample to be quite dilute, so that the lifetime of the three-component gas is much longer than the time required to transfer the atoms. Additionally, for our measurements the gas has to be cold enough that temperature effects do not wash out the resonances \cite{jose_temp}. At the same time we cannot allow the gas to become quantum degenerate, as this would lead to the formation of a molecular BEC, which would spatially separate from the rest of the cloud and make a quantitative determination of the loss rates much more difficult. To simultaneously fulfill all these conditions we use two accousto-optical modulators to adiabatically increase the radial extension of the trap by a factor of four by creating a time-averaged potential. This leads to an adiabatic decrease of the temperature to its final value of $60 \pm 15 \,$nK  and lowers the peak density by about an order of magnitude to a value of roughly $1.5 \times 10^{11} (2 \times 10^{11})\, \rm{atoms\,(molecules)}/ \rm{cm}^3$. The final trap frequencies are $2\pi \times 102(5)$ and $2\pi \times 112(5)$ Hz in the radial and $2\pi \times 15(1)$ Hz in the axial direction. After this reduction in density we can ramp to the magnetic field of interest and drive the free atoms in state $\left|2\right\rangle$ to state $\left|1\right\rangle$ within $500\,\rm{\mu s}$ using a rapid adiabatic passage. This preparation scheme is unique to fermionic systems, as two-component Fermi gases are very stable against inelastic decay close to Feshbach resonances \cite{petrov_pra}. For smaller scattering lengths, losses through $\left|2\right\rangle$-$\left|23\right\rangle$ atom-dimer and $\left|23\right\rangle$-$\left|23\right\rangle$ dimer-dimer collisions become more important, but are still low enough that we lose only a small amount of particles during the final field ramp.

\begin{figure} [tb]
\centering	\includegraphics [width= 8cm] {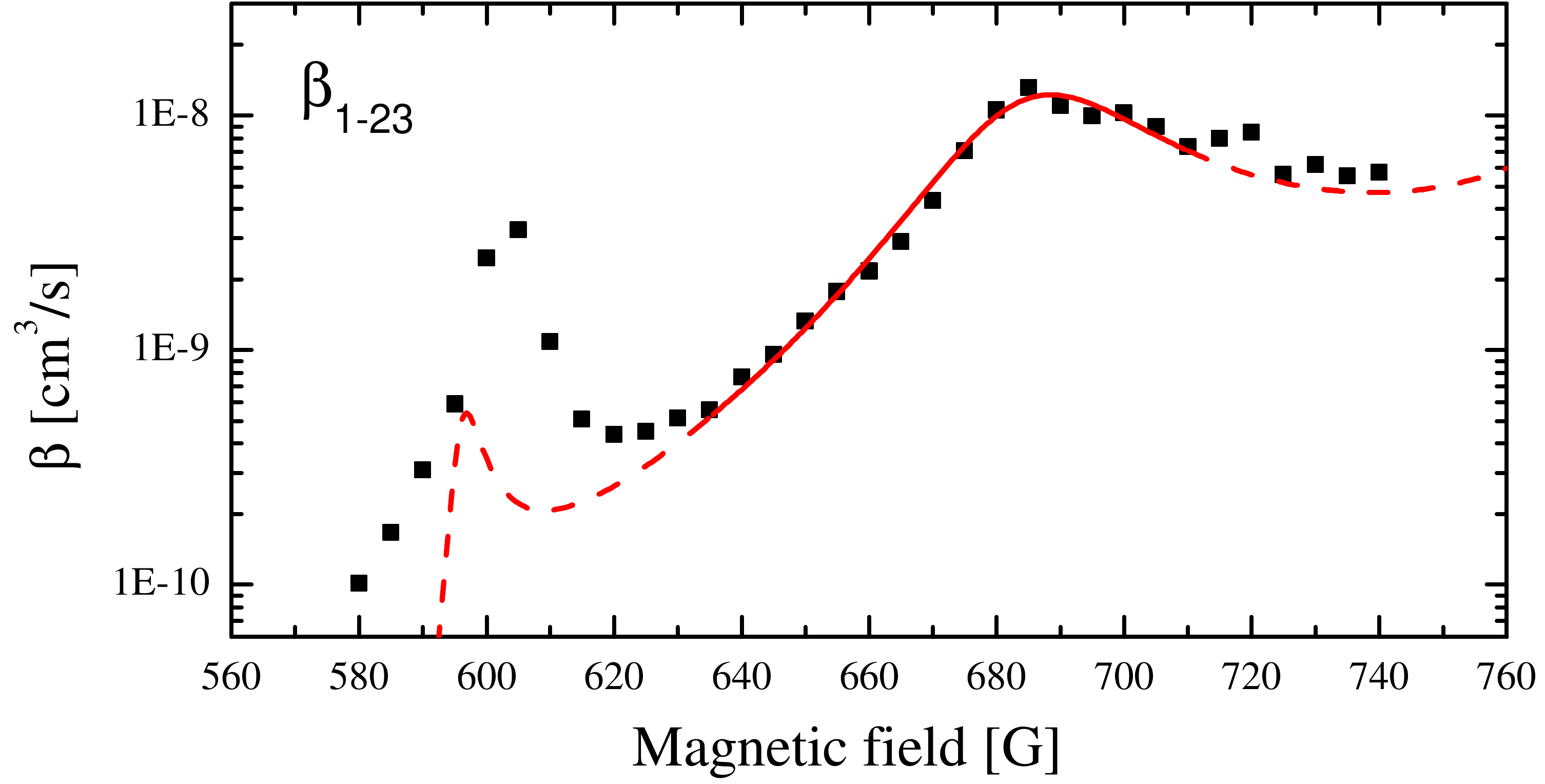}  
	\caption{ Two-body loss rate for a mixture of state $\left|1\right\rangle$ atoms and $\left|23\right\rangle$ dimers vs. magnetic field. The red line is a fit according to eq. \ref{beta_braaten_hammer}, the solid part of the line marks the magnetic field range included in the fit. 	\label{fig:beta_1-23}}
\end{figure}

After preparing the $\left|1\right\rangle$-$\left|23\right\rangle$ mixture we record the number of remaining atoms over 120\,ms with in-situ absorption imaging. Due to the low density and temperature of the sample we cannot use time-of-flight imaging to follow the temperature evolution of the cloud. Additionally, for high loss rates the decay happens much faster than thermalization, so that the sample falls out of thermal equilibrium. As a quantitative treatment of these effects is prohibitively complex, we assume the sample to have a Gaussian density distribution and fit only the initial decay, where the effects of heating and anti-evaporation should be only a small correction \cite{note_beta_fit}.
If we neglect dimer-dimer collisions the two-body loss is described by the following differential equation \cite{note_dimer_dimer}:
\begin{equation}
\dot{N}_A =  - \beta \, \int n_A(\textbf{r}) \, n_D(\textbf{r}) \, d\textbf{r}^3,
\end{equation}
where $\beta$ is the two-body loss coefficient, $N_A$ is the atom number and $n_A(\textbf{r})$ ($n_D(\textbf{r})$) is the atomic (molecular) density. This equation has the analytical solution
\begin{equation}
{N_A}(t) =  \frac{(N_{D0} - N_{A0})}{\frac{N_{D0}}{N_{A0}} e^{(N_{D0} - N_{A0}) \gamma t} - 1}
\label{loss_curve} 
\end{equation}
with $\gamma = \beta / ( \sqrt{27}\pi^{\frac{3}{2}} \sigma_x \sigma_y \sigma_z  )$, where $\sigma_x$, $\sigma_y$ and  $\sigma_z$ are the $1/\sqrt e$ radii of the atom cloud and $N_{A0}$ ($N_{D0}$) is the initial atom (molecule) number at $t=0$.
Using this model we can extract the atom-dimer loss rate $\beta_{1-23}$ from the loss curves. The results are plotted in fig. \ref{fig:beta_1-23}. 

In the magnetic field region below $730$\,G the $\left|23\right\rangle$ dimer is the lowest-lying weakly bound dimer state. Therefore the only possible atom-dimer loss process is relaxation into deeply bound dimer states. We observe two distinct maxima of the loss rate, which we attribute to the two predicted trimer states crossing the $\left|1\right\rangle$-$\left|23\right\rangle$ atom-dimer threshold.  According to \cite{braaten_6li} the relaxation rate around the crossing of the second trimer is
\begin{equation}
\beta_{1-23} =  \frac{6.32 C \sinh(2\eta_*)}{\sin^2(s_0 \ln(a_{23}/a_*^+)) + \sinh^2\eta_*} \frac{\hbar a_{23}}{m},
\label{beta_braaten_hammer} 
\end{equation}
in the zero-energy and zero-range limit, where the position $a_*^+$ and width $\eta_*$ of the resonance are free parameters. 
To account for systematic shifts in our data due to errors in our determination of particle density, which we estimate to be $\sim 80\%$ due to uncertainties in magnification, particle number determination and trap frequencies, we have included an additional normalization constant C in our fit.  We only fit the data for the region where $a_{23} > 5 \times r_0$ and $E_{23} - E_{12} \gg k_BT$ (solid line in fig.\ref{fig:beta_1-23}). We find good quantitative agreement between the fit and our data in this region, with fit parameters $a_*^+ = 1076\,(1042,1110) $\,$a_0$, $\eta_* = 0.34\,(0.29,0.39)$ and $C=0.52\,(0.47,0.57)$, where the values in parantheses give the $95\%$ confidence bounds of the fit. This corresponds to a resonance position of $B_* = 685 \pm 2$\,G. One should note that the highest measured value of $\beta_{1-23}$ of $1.3 \times 10^{-8}$\,cm$^3$/s is only a factor of three lower than the unitary limit of $4 \times 10^{-8}$\,cm$^3$/s for our temperature \cite{jose_limit}, which can cause additional systematic errors for $C$ and $\eta_*$.
For the lower resonance eq. \ref{beta_braaten_hammer} is not valid any more, so we fit its position independently. A Gaussian fit to the peak yields a position of $603 \pm 2$\,G, which corresponds to a scattering length of $a_{23} \sim 90$\,$\rm{a}_0$, so the universal description is not expected to be quantitatively accurate in this region.

These results qualitatively confirm the predictions about the spectrum of Efimov states made in \cite{braaten_6li}, yet the three-body parameter determined from the upper atom-dimer resonance differs by about $30 \%$ from the one observed at $895$\,G  \cite{o'hara_oben}. This suggests a magnetic field dependence of the three-body parameter which cannot be described by universal theory.

\begin{figure} [tb]
\centering	\includegraphics [width= 8cm] {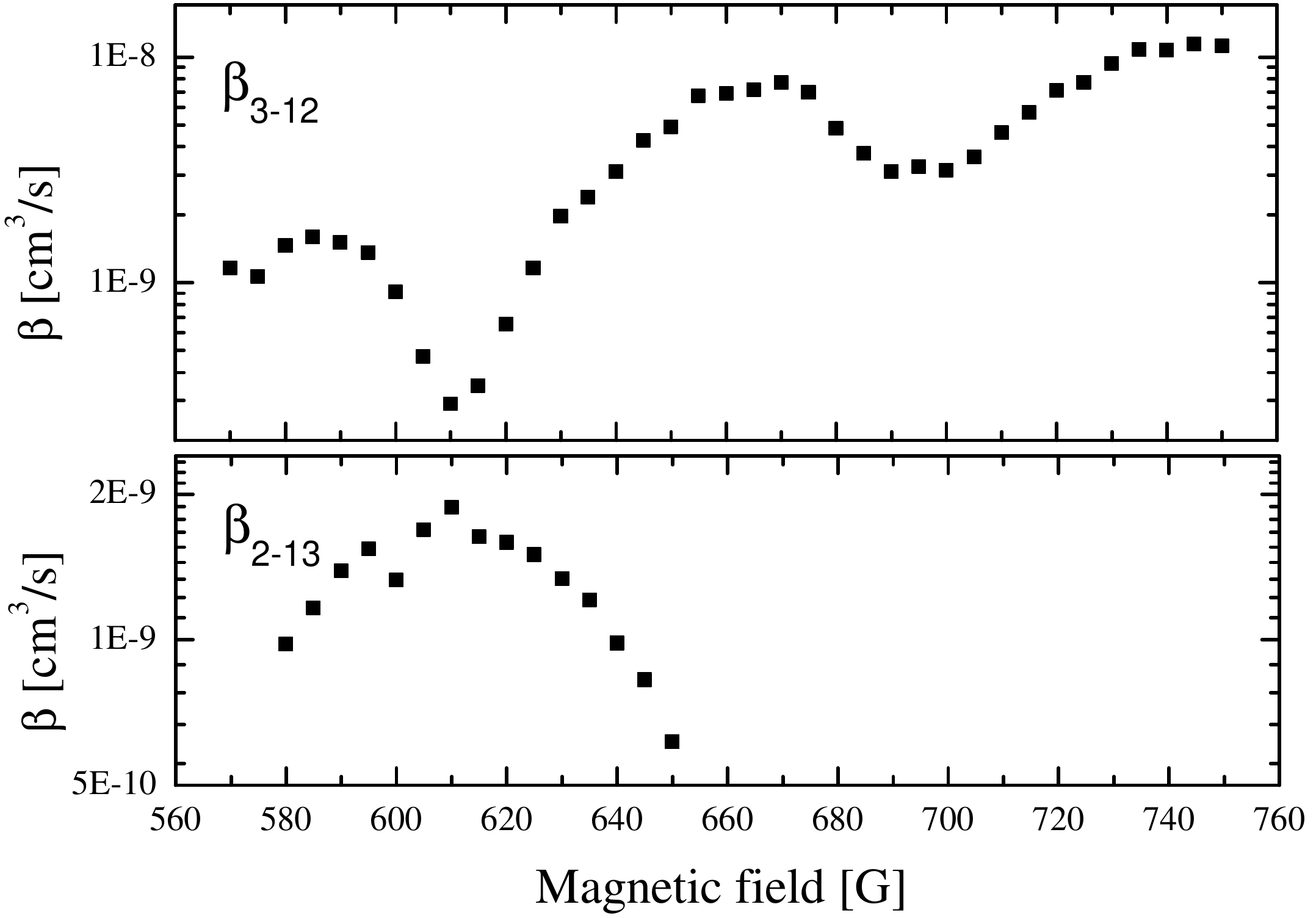}  
	\caption{Measured two-body loss rate for the $\left|3\right\rangle$-$\left|12\right\rangle$  and $\left|2\right\rangle$-$\left|13\right\rangle$ atom-dimer mixtures vs. magnetic field. 
	\label{fig:beta_other}
	}
\end{figure}

To study the $\left|3\right\rangle$-$\left|12\right\rangle$ and $\left|2\right\rangle$-$\left|13\right\rangle$ mixtures we use the same techniques as described above.
For the $\left|3\right\rangle$-$\left|12\right\rangle$ mixture the decay rate $\beta_{3-12}$ results from a combination of two inelastic processes: Relaxation of $\left|12\right\rangle$ molecules into deep dimers, and recombination events of the form $\left|3\right\rangle$-$\left|12\right\rangle \rightarrow \left|1\right\rangle$-$\left|23\right\rangle$. Similar exchange processes have already been observed in ultracold Cesium \cite{innsbruck_exchange}. For fields below $720$\,G, where the trap depth is much lower than the difference between the $\left|12\right\rangle$ and $\left|23\right\rangle$ binding energies both processes lead to trap loss, so we cannot distinguish between the two. For fields above the crossing of $a_{12}$ and $a_{23}$ at $730$\,G the $\left|12\right\rangle$-dimer is the lowest-lying shallow dimer and relaxation into deep dimers is the only loss process.
The results for $\beta_{3-12}$ are shown in fig \ref{fig:beta_other}.

While we observe an overall increase of the loss coefficients for higher magnetic fields, we find two pronounced minima of $\beta_{3-12}$ at $610$\,G and $695$\,G where the loss rate is strongly suppressed.  The rates for relaxation events and exchange reactions close to overlapping Feshbach resonances have been calculated by J. P. D'Incao et al. for the special case  $\left|a_{13}\right| \gg a_{12} \gg a_{23}$ \cite{jose_overlapping}. As the condition $a_{12} \gg a_{23}$ is not fulfilled for the case of $^6\rm{Li}$ we cannot use the formulas given in \cite{jose_overlapping} to fit our data, but the presence of two minima agrees well with the theoretical prediction that a trimer state can cause interference minima in the rate of exchange reactions.
Therefore we believe that the observed minima of $\beta_{3-12}$ at $610$\,G and $695$\,G are caused by interference minima in the rate of $\left|3\right\rangle$-$\left|12\right\rangle \rightarrow \left|1\right\rangle$-$\left|23\right\rangle$ exchange reactions, which are visible on top of a background of relaxation events \cite{note_exchange}. 
The fact that both minima appear at slightly higher fields than the resonant enhancements of $\beta_{1-23}$ suggests a connection between these features. For the features at $685$\,G and $695$\,G this connection should be universal, i.e. the positions of the features should be determined by the same three-body parameter.

For the $\left|2\right\rangle$-$\left|13\right\rangle$ mixture the decay is a combination of $\left|2\right\rangle$-$\left|13\right\rangle \rightarrow \left|1\right\rangle$-$\left|23\right\rangle$ and $\left|2\right\rangle$-$\left|13\right\rangle \rightarrow \left|3\right\rangle$-$\left|12\right\rangle$ recombination and relaxation into deep dimers. We have measured the decay rate for fields between $580$\,G and $650$\,G and find that the observed decay rate shows no prominent features. However, we cannot exclude the possibility that there are features which are masked by higher loss in another channel.

In summary, we have prepared three different atom-dimer mixtures of $^6\rm{Li}$ and studied their decay by inelastic atom-dimer collisions. We have pinned down the crossings of the two predicted universal trimers with the $\left|1\right\rangle$-$\left|23\right\rangle$ continuum by observing two loss resonances in the $\left|1\right\rangle$-$\left|23\right\rangle$ atom-dimer scattering. This completes the observation of the spectrum of Efimov states in $^6\rm{Li}$. 
In the $\left|3\right\rangle$-$\left|12\right\rangle$ mixture we observe two interference minima where the rate of molecular exchange reactions is strongly suppressed by the presence of lower lying trimer states. This leads to a significant increase in the lifetime of the gas, which can be exploited in future experiments. 
The knowledge of the few-body physics and the experimental techniques developed in this work are a vital step forward on the way to studying many-body physics of three-component Fermi gases.

We thank J. Ullrich and his group for their generous support. We thank E. Braaten, J. P. D'Incao, H. W. Hammer and D. S. Petrov for valuable discussions.

After submission of this paper we became aware of similar studies by S. Nakajima et al. \cite{naidon_atomdimer}.

\bibliography{paper}

\begin{thebibliography}{32}
\bibitem[{not({\natexlab{a}})}]{note_free_minority}
\bibinfo{note}{{We have verified the fraction of free minority atoms to be $< 5
  \%$ by imaging at small scattering length where atoms bound in dimers are not
  visible.}}
\bibitem[{not({\natexlab{b}})}]{note_beta_fit}
\bibinfo{note}{{We have fitted the data up to a loss fraction of 0.3, 0.5 and 1
  and have found the variation between fitting to a loss fraction of 0.3 or 0.5
  to be $\sim 10 \%$ }}.
\bibitem[{not({\natexlab{c}})}]{note_dimer_dimer}
\bibinfo{note}{{We have independently measured the dimer-dimer loss rates and
  found $\beta_{DD} < \beta_{AD}$ except for the
  $\left|1\right\rangle$-$\left|23\right\rangle$ mixture at fields below
  595\,G.}}
\bibitem[{not({\natexlab{d}})}]{note_exchange}
\bibinfo{note}{{Direct observation of the reaction products is only possible if
  they remain trapped, which requires that the kinetic energy released in the
  reaction is smaller than the trap depth. Unfortunately in our case this
  condition is only fulfilled in a region of the magnetic field where the
  exchange reactions are strongly suppressed; consequently we cannot directly
  observe the reaction products.}}
\end{thebibliography}

\end{document}